\documentclass{PoS}
\pdfoutput=1

\usepackage[utf8]{inputenc}
\usepackage{amsmath,amssymb,amstext}
\usepackage{subfigure}

\newcommand{\iu}{\mathrm{i}}

\newcommand{\unit}[1]{~\text{#1}}

\title{Electromagnetic corrections to pseudoscalar decay constants}

\ShortTitle{Electromagnetic corrections to pseudoscalar decay constants}

\author{\speaker{Benjamin Gläßle},%
		\, Gunnar S. Bali\\
        Institut für Theoretische Physik, Universität Regensburg \\
		93040 Regensburg, Germany\\
        E-mail: \email{benjamin.glaessle@physik.uni-regensburg.de} \\
		\qquad\quad\,	\email{gunnar.bali@physik.uni-regensburg.de}\\
		}

\author{\rm \centering (QCDSF Collaboration)}

\abstract{The effects of electromagnetic interactions on pseudoscalar decay constants are investigated. Using a compact QED and QCD action we are able to resolve differences of about $0.1\unit{MeV}$. We obtain the preliminary results $f_{\pi^0}-f_{\pi^\pm}=0.09(3)\unit{MeV}$ and $f_{D^0}-f_{D^\pm}=0.79(11)\unit{MeV}$ for light and charmed pseudoscalar decay constants on a $N_\text{f}=2$ nonperturbatively improved Sheikholeslami-Wohlert ensemble.}
\FullConference{XXIX International Symposium on Lattice Field Theory \\
		 July 10 – 16 2011\\
		 Squaw Valley, Lake Tahoe, California}

\begin{document}

\section{Introduction}
\noindent
With increased precision of lattice calculations it becomes necessary to investigate and
to remove the approximations that are often made. One such approximation is the use of unphysically heavy sea quark masses, another one is the omission of electromagnetic effects. From the mass differences of charged and uncharged pions or between the nuclei one would expect these to be of the order of one to a few MeV.

Recent lattice calculations \cite{precDs1,precDs2} of pseudoscalar $D_s$ decay constants have reached an accuracy of one percent, a regime where electromagnetic effects may become significant. The systematic errors stated
in these calculations include an estimate of QED effects that is based on electromagnetic shifts of
the $D_s$ meson mass \cite{precDs2}.

The methods initially introduced in \cite{Duncan96} make it possible to include QED effects on the lattice explicitly and thereby enable us to differentiate between isospin breaking through charge and different up and down quark masses. These techniques have been successfully applied in calculations of the light hadron spectrum \cite{Duncan96,Blum2010,Portelli2010} and used to estimate the $u$,$d$ mass difference. We
deviate from these references, using a compact QED action, to
investigate electromagnetic effects on pseudoscalar decay constants.

\section{Methods}
\noindent
Electromagnetic effects on the lattice are included by multiplying the QCD SU(3)-links with QED U(1)-links, which are given by $U_{QED,\mu}(x)=\exp\left( \iu e B_{\mu}(x) \right)$, where $e$ is the charge of the corresponding quark. The calculation of fully unquenched SU(3)$\times$U(1)$\rightarrow$U(3) configurations is practically unfeasible and unnecessary since sea quark charge effects are suppressed by an
additional factor of $\alpha_{QED}\approx 1/137$. Instead, we use quenched QED configurations
and multiply these with $N_{\text{f}}=2$ QCD configurations. This is correct up to $\mathcal{O}(\alpha_{QED})$.

\begin{figure}
\centering
\includegraphics[width=0.88\textwidth]{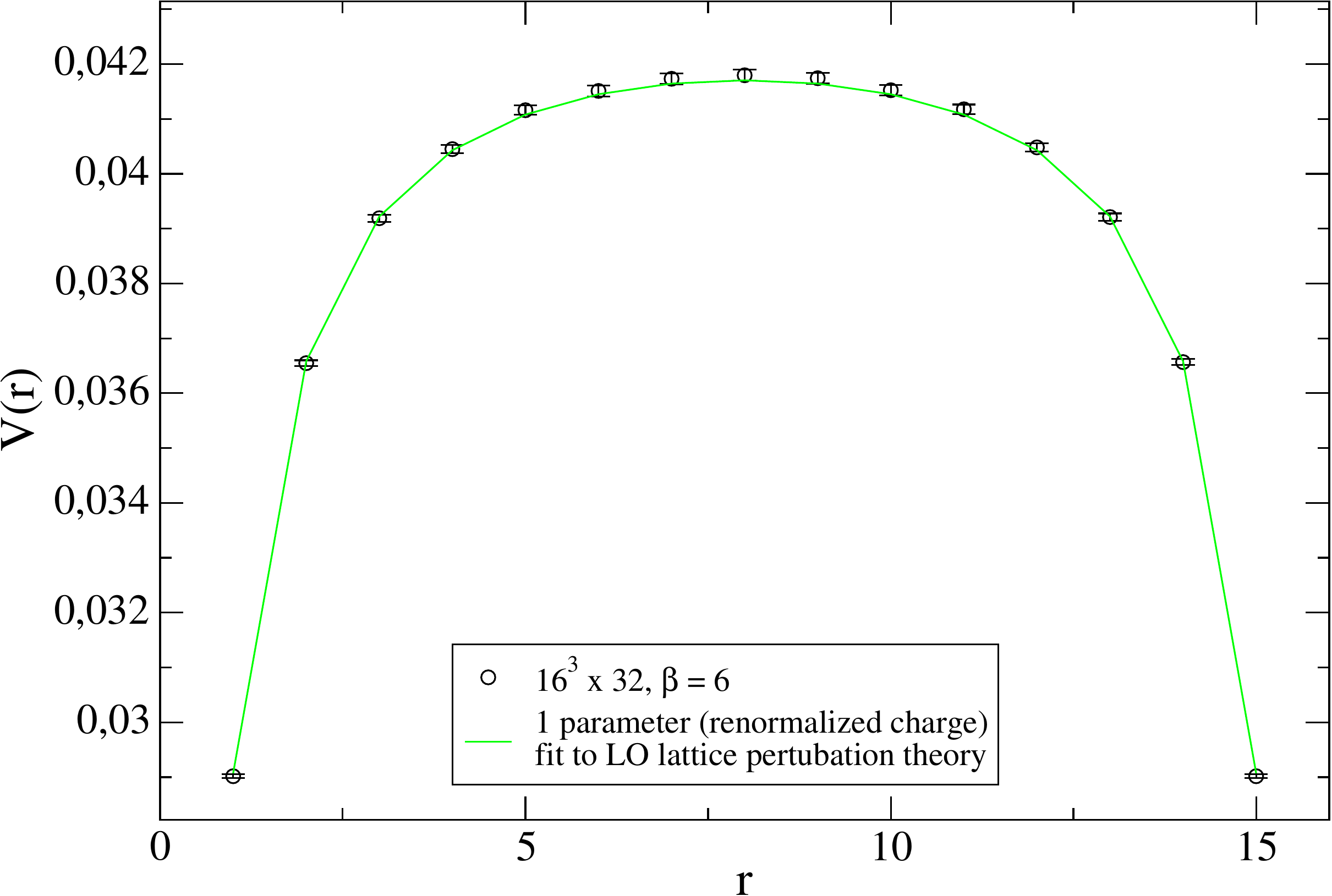}
\caption{Coulomb potential: measured data and prediction of lattice pertubation theory.}
\label{fig:vr}
\end{figure}

Instead of using a noncompact action as in previous lattice studies of electromagnetic effects \cite{Duncan96,Blum2010,Portelli2010}, we employ the compact Wilson gauge plaquette action, so we
can deal with these degrees of freedom on the same footing as with the SU(3) gauge links.
Starting from a cold (unit link) lattice, we employ the heatbath algorithm to obtain
the $B_{\mu}(x)$ phases by rescaling random variables $\theta_x$ that are distributed
according to $P(\theta_x) \sim \exp( w_x \cos \theta_x)$ where the weights $w_x$ depend on the coupling and on the local staple. We follow the procedure described in \cite{Hattori92}. The lattice is divided into 4-cubes with side length 2 and our algorithm loops over single sites of every 4-cube simultaneously. We find integrated autocorrelation times of the plaquette of $\mathcal{O}(1)$. Nonetheless, to be on
the safe side, only every 200$^{\text{th}}$ QED configuration is used. The first 3000 configurations are
discarded such that the ensemble is sufficiently thermalized. The improvement coefficient of the fermionic
QCD Sheikholeslami-Wohlert action $c_{sw,QCD}$ has been determined nonperturbatively.
Since the QED interactions are quenched this coefficient will not be affected by them (and in the
unquenched case it would receive an $\mathcal{O}(\alpha_{QED})$ shift only).
To be consistent to leading order in the QED coupling we set $c_{sw,QED}=1$.

We treat the U(1)-links as a background and set their coupling to the physical value.
Because quark charges are multiples of one third of the positron charge $e_{p^+}$, we rescale the coupling such that the smallest (nonzero) charge is $e = \frac{e_{p^+}}{3}$. This means that $\beta^R\approx 3^2/(4\pi\alpha_{QED})\approx 98$.
This can then easily be rescaled to $e_q\,e$ with $e_q=\{-1,\pm2\}$ to calculate propagators of other charges.
We use the bare value $\beta=99$ (that corresponds to the above renormalized coupling)
for the generation of
the U(1)-links. The previously mentioned heatbath algorithm deals well with the resulting,
very narrow distribution $P(\theta_x)$. Since we are using a compact formulation of the
action there are unphysical 4-point and higher order vertices, which renormalize the coupling.
This is no problem since this effect is purely multiplicative and does not introduce any
running. This means that $\beta^R=Z\beta$, with a renormalization factor
$Z=1+\mathcal{O}(1/\beta)$. We determine this factor nonperturbatively by
comparison between the static lattice QED potential and the analytic perturbative expectation
that, without Fermions, should be exact in noncompact QED. This is visualized in Fig.~\ref{fig:vr}
for $\beta=6$.
From the multiplicative constant that is fitted, we determine $Z\approx 0.99$ at our
$\beta$-value.

To suppress large finite-size effects for correlation functions of charged particles, that
are not gauge invariant with respect to U(1), zero modes of the electromagnetic field need to be
subtracted \cite{QEDFiniteSize08}. This is achieved by a global gauge transformation \cite{Bogolubsky99}.
In addition, we fix the U(1)-links to Landau gauge, to improve the signal.

Including electromagnetic effects additively renormalizes the mass of charged quarks in the
Wilson formulation. We choose to set the quark masses in such a way that their renormalized
values are (approximately) the same. The strange $\kappa$ is fixed by tuning the mass of the
hypothetical $s\bar{s}$ pseudoscalar to the experimental value,
\begin{align}
m^2_{K^0} + m^2_{K^+} - m^2_{\pi^\pm} \simeq (690\unit{MeV})^2 = m^2_{s\bar{s}}.
\end{align}
This is done for the charges $e_\text{strange}=\{0,\pm1,\pm2\}$. The charm mass parameter ($\kappa_{\text{charm}}$) is only determined for $e_\text{charm}=\{0,\pm2\}$.
It is fixed by tuning $D_s=c\bar{s}$ to its physical mass for uncharged charm and strange quarks as well as for $e_\text{charm}=2$ and $e_\text{strange}=-1$. To check the tuning of the charm $\kappa$s we also calculated the mass combination $M_\text{charm} = \frac{1}{4}\left(3 m_{J/\Psi}+m_{\eta_c}\right)$.
\begin{center}
\begin{tabular}{|c|c|}
\hline
$e_\text{charm}$ & $M_\text{charm}$ \\
\hline
exp. & $3.0678(3) \unit{GeV}$ \\
0 & $3.002(2) \unit{GeV}$ \\
$\pm2$ & $3.024(2) \unit{GeV}$ \\
\hline
\end{tabular}
\end{center}
The difference between the 3.024~GeV and the experimental value may be attributed to the unphysical
sea quark content and finite size effects. However, the difference between our two
calculations shows that there is some inconsistency in the tuning of the charm quark mass between
the two procedures of about 10~MeV. This is no surprise since the total charges of the two mesons differ. We regard tuning of the $\bar{c}c$ combination
the cleaner procedure since this does not depend on the strange quark mass and we will implement this in the future.

We also simulate 2 lighter quarks of all five charges to enable extrapolations to physical light quark masses
for pions and for the $D^{0,\pm}$. We tune the symmetric and therefore uncharged combinations $l\bar{l}$ for all charges $e_l=\{0,\pm1,\pm2\}$ to the same values $m_{l\bar{l}}^2$.                                                                                                                                                                                                                                                                                                                                            To reduce the noise we average over $\pm B_{\mu}$ \cite{Blum2010} which is equivalent to averaging over $\pi^\pm$. Our $\kappa$-values are listed in the table below.
\begin{center}
\begin{tabular}{|c|llll|}
\hline
$e_q$ & $\kappa_{l_1}$ & $\kappa_{l_2}$ & $\kappa_\text{strange}$ & $\kappa_\text{charm}$ \\
\hline
0 & 0.13629 &  0.136013 & 0.135676 & 0.123019 \\
$\pm1$ & 0.136337 & 0.13606 & 0.135722 & - \\
$\pm2$ & 0.136477 & 0.136199 & 0.135861 & 0.123086 \\
\hline
\end{tabular}
\end{center}

We use spin-explicit, complex $Z_2$ random wall sources \cite{OneEndSEM2008}. This enables us to average over the spatial volume. We use 3 noise sources per configuration and analyze a total of 200
$N_\text{f}=2$ nonperturbatively improved Sheikholeslami-Wohlert configurations \cite{QCDSFgen} at $\beta=5.29$, $\kappa=0.1355$ on a $24^348$ volume. The sea pion mass reads $m_{\pi,sea}\simeq 750\unit{MeV}$ and
we use $a=0.086\unit{fm}$ as our lattice spacing. All 2-point functions are calculated for a local and a Wuppertal smeared sink. Wuppertal smearing was done using the product of separately APE smeared QCD- and QED-links.
The decay constants are extracted from $\pi\pi$ and $\pi A_4$ correlators and have been improved and renormalized according to \cite{Gockeler:1997fn}, neglecting $\alpha_{QED}$ corrections since we consistently work at
$\mathcal{O}(\alpha_{QED})$. We are using the Chroma software system \cite{CiteChroma} which has been modified to support QCD+QED calculations.

\section{Analysis}
\noindent
The available data can also be used to extract the up/down quark masses and the pion mass splitting, which is done to verify the correctness of the approach. To do this all pseudoscalar mass data must be fitted to a modified chiral expression which takes QED into account. Instead of using the simple formula \cite{Duncan96}
\begin{align}
  m_{PS}^2 = A_0 Q^2 + \left(B_0 + B_1 Q^2\right) \left( m_q + m_{\bar{q}} \right),
  \label{eq:totalChiPT}
\end{align}
where $Q=e_q+e_{\bar{q}}$, to fit charged and uncharged data, we alternatively subtract the uncharged mesons (with the parameters obtained by the $\kappa_c$ fit) and fit this difference to 
\begin{align}
  \Delta m_{PS}^2 = A_0 Q^2 + B_1 Q^2 \left( m_q + m_{\bar{q}} \right).
\end{align}
Once the fit parameters $A_0,B_0$ and $B_1$ are determined the experimental values of $m_{K^{0,\pm}}$ and $m_{\pi^\pm}$ can be used to determine the masses of up, down and strange quarks. The physical $\pi$ mass difference is retrieved by reinserting all this information into eq. (\ref{eq:totalChiPT}).
The result is in good agreement with experimental value (Tab. \ref{tab:qmassdiff}) despite a relatively large $\chi^2/$dof$=4.5$. We plan to address this in future studies, for example by including higher order terms in the chiral fit.
\begin{table}
\begin{center}
\begin{tabular}{|c|cc|}
\hline
 &  lattice result & experimental \\
\hline
$m_d/m_u$ & $1.80(4)$ & - \\
$\Delta m_{\pi}$ &  $4.4(8)\unit{MeV}$ & $4.5936(5)\unit{MeV}$ \\
\hline
\end{tabular}
\end{center}
\caption{Quark masses and the pion mass differences. The errors are only statistical.}
\label{tab:qmassdiff}
\end{table}

The calculated decay constants differ by about 10~\% from the
experimental results, which is not surprising since this is only a partially quenched study on an ensemble with a relatively large sea quark mass, at one lattice spacing. By computing differences between differently charged pseudoscalars at the same mass scale we are able to extract 
the QED contributions. However, the decay constant will depend on the mass and this will depend on the
charge of the particle. To disentangle these two effects we perform the comparison, matching the
uncharged squared mesons masses $m^2_{PS,light}(Q=0)$ of each of the two constituent quarks. 
As an example consider the positively charged Kaon, which consists of a light quark $l$ with charge $2$ and a strange antiquark with charge $1$: The corresponding scale would be
$m^2_{PS,light}(Q=0)=\frac{1}{2}( m^2_{l\bar{l}} + m^2_{s\bar{s}} )$. This is equivalent to using
the average light quark mass, but more correlations between the data remain, because the
$\kappa_{crit}(e)$-values do not have to be determined. Preserving these correlations is necessary
if differences significantly smaller than the noise of the absolute signals are extracted.
In the case of the charm quark the Gell-Mann-Oakes-Renner relation does not apply anymore
so that we cannot perform chiral fits and moreover, it would not be sensible to match
squared pseudoscalar masses. The charm quark matching has not been done as yet so that in this case we cannot
compare the decay of the physical $D_s$ meson to that of a $D_s$ with electrically neutral
quarks. However, we can compare the decay constants obtained for combinations of the
$e_{\text{charm}}=2$ charm quark with differently charged light quarks.

In the case of light mesons we restrict ourselves to (approximately) equal valence
quark masses and compare the other decay constants with $f_{\pi^0} = \frac{1}{2}(f_{u\bar{u}} + f_{d\bar{d}})$, where $u$ and $d$ quark masses are approximately equal and have the appropriate (physical) charges. Neglecting disconnected loops for $\pi^0$ is correct to $\mathcal{O}(\alpha_{QED}^2)$. The effects of isospin breaking by different $u$ and $d$ quark masses have not been investigated. Under these conditions we are able to extract the difference $f_{\pi^0}-f_{\pi^\pm} = 0.09(3)\unit{MeV}$ from a fit to correlated data. This estimate is already more  precise than the experimental value.

\begin{figure}
\centering
\subfigure[Pions]{
  \label{fig:fll}
  \includegraphics[width=0.47\textwidth]{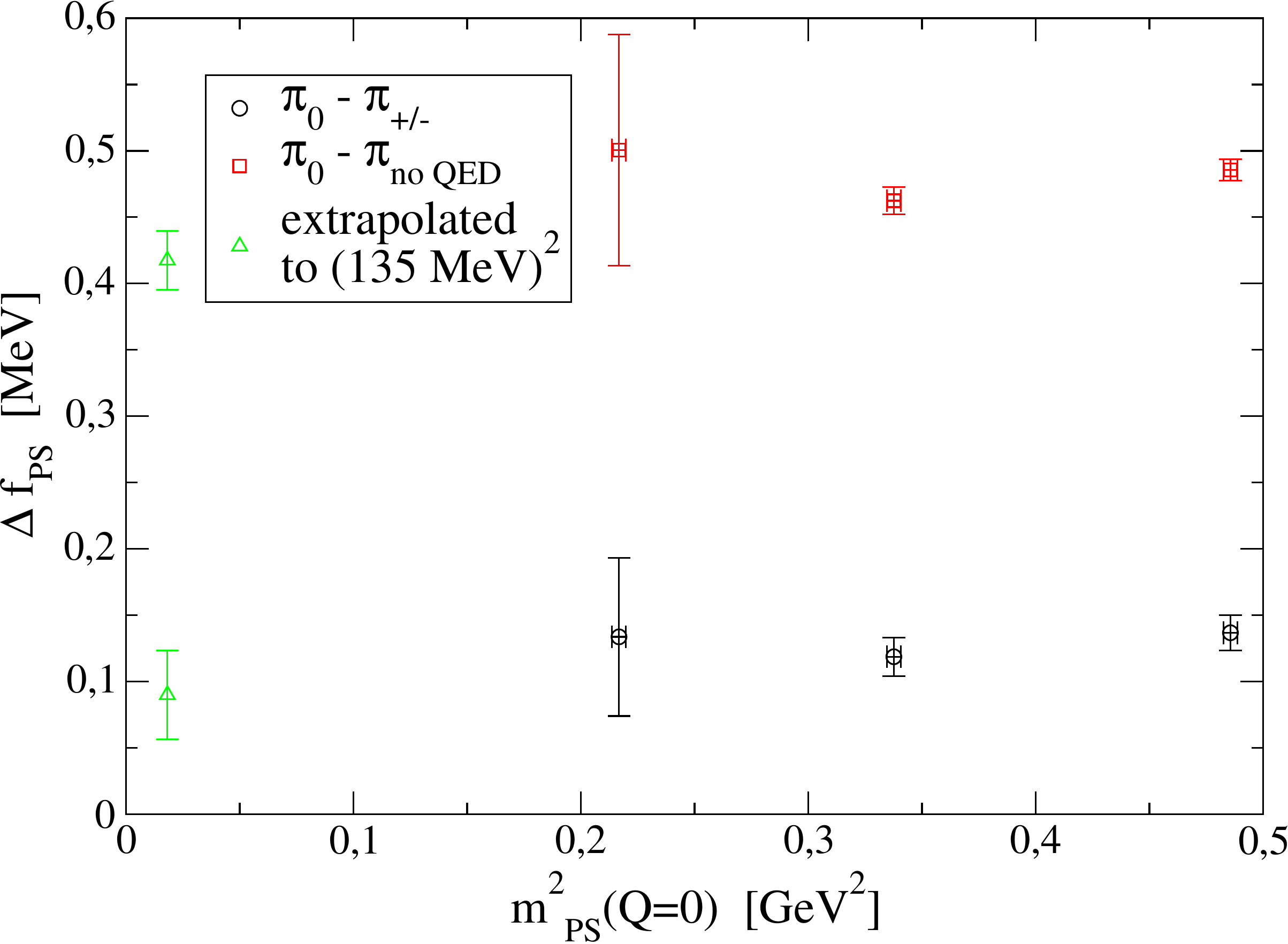}
}
\subfigure[charmed pseudo scalars]{
  \label{fig:fhl}
  \includegraphics[width=0.47\textwidth]{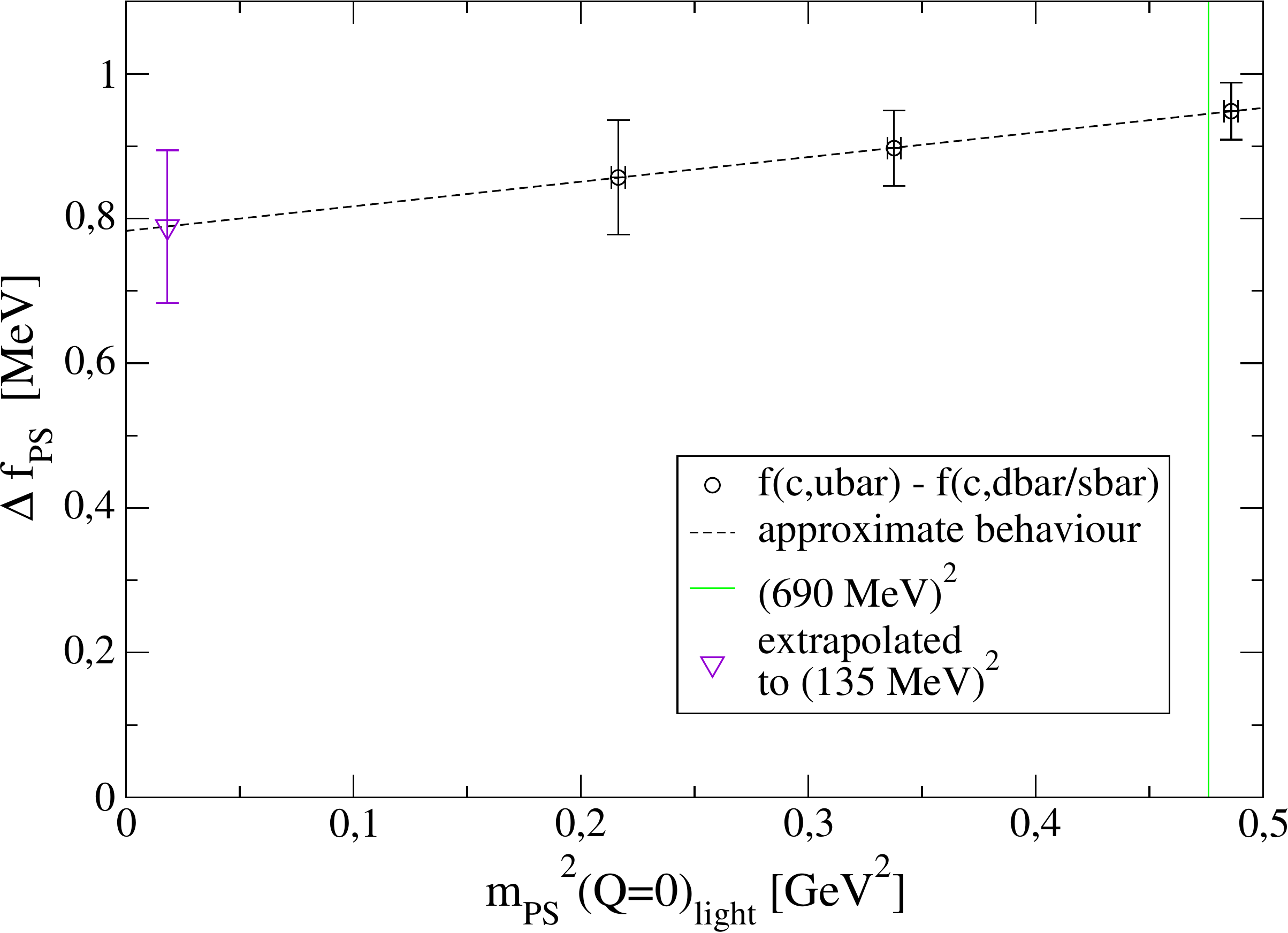}
}
\caption{Differences of the decay constants between uncharged and charged mesons. All errors are only statistical.}
\end{figure}

In the case of the $D$ mesons we are limited to the difference between the $D^0$ and charged $D^\pm$: $f_{D^0}-f_{D^\pm}=0.79(11)\unit{MeV}$. If the light mass is set to the strange mass, we can define the QED contribution as the difference between the charged $D_s$ and a hypothetical, uncharged $D_s^0$ meson, where the antistrange has charge $e_{\bar{s}}=-2$: $f_{D_s^0}-f_{D_s}=0.95(4)\unit{MeV}$. The order of magnitude and the sign of our result agree with naive expectations.

\section{Conclusion}
\noindent
We are able to resolve differences between decay constants of differently charged mesons
of about 0.1~MeV. We determine the electromagnetic effect on $D$, $D_s$ and $\pi$ decay constants.
The difference between charge and uncharged $\pi$ decay constants is more precise than experimental
results. We find $f_{D_s}$ to decrease by about 1~MeV, due to electromagnetic effects.

\section*{Acknowledgments}
\noindent
We thank 
J. Najjar, S. Collins and L. Castagnini
for their help.
This work was supported by the GSI Hochschulprogramm (RSCHAE), the European Union
under Grant Agreement number 238353 (ITN STRONGnet)
and by the Deutsche
Forschungsgemeinschaft SFB/Transregio 55.
Computations were performed on Regensburg's Athene HPC Cluster.

\end{document}